# Diverse response of surface ozone to COVID-19 lockdown in China

Yiming Liu[a,*#], Tao Wang[a,*], Trissevgeni Stavrakou[b], Nellie Elguindi[c], Thierno Doumbia[c], Claire Granier[c,d], Idir Bouarar[e], Benjamin Gaubert[f], Guy P. Brasseur[a,e,f]

[a]Department of Civil and Environmental Engineering, The Hong Kong Polytechnic University, Hong Kong, China;

[b]Royal Belgian Institute for Space Aeronomy, Brussels, Belgium;

[c]Laboratoire d'Aérologie, Toulouse, France;

[d]NOAA Chemical Sciences Laboratory and CIRES/University of Colorado, Boulder, CO, USA;

[e]Environmental Modeling Group, Max Planck Institute for Meteorology, Hamburg, Germany;

[f]Atmospheric Chemistry Observations and Modeling Laboratory, National Center for Atmospheric Research, Boulder, CO, USA;

[*]*Correspondence to*: Tao Wang (cetwang@polyu.edu.hk) and Yiming Liu (liuym88@mail.sysu.edu.cn)

[#]Now in School of Atmospheric Science, Sun Yat-sen University, Zhuhai, China

**Abstract:** Ozone ($O_3$) is a key oxidant and pollutant in the lower atmosphere. Significant increases in surface $O_3$ have been reported in many cities during the COVID-19 lockdown. Here we conduct comprehensive observation and modeling analyses of surface $O_3$ across China for periods before and during the lockdown. We find that daytime $O_3$ decreased in the subtropical south, in contrast to increases in most other regions. Meteorological changes and emission reductions both contributed to the $O_3$ changes, with a larger impact from the former especially in central China. The plunge in nitrogen oxide ($NO_x$) emission contributed to $O_3$ increases in populated regions, whereas the reduction in volatile organic compounds (VOC) contributed to $O_3$ decreases across the country. Due to a decreasing level of $NO_x$ saturation from north to south, the emission reduction in $NO_x$ (46%) and VOC (32%) contributed to net $O_3$ increases in north China; the opposite effects of $NO_x$ decrease (49%) and VOC decrease (24%) balanced out in central China, whereas the comparable decreases (45-55%) in these two precursors contributed to net $O_3$ declines in south China. Our study highlights the complex dependence of $O_3$ on its precursors and the importance of meteorology in the short-term $O_3$ variability.

# 1 Introduction

The outbreak of coronavirus disease 2019 (COVID-19) has severely threatened public health worldwide, leading to millions of deaths(WHO, 2020). China, where the first case of COVID-19 was reported in the city of Wuhan, imposed country-wide measures from 23 January to 13 February 2020 to prevent the spread of the disease, including social distancing, teleworking, and closure of non-essential businesses (Chinazzi et al., 2020; Li et al., 2020). These restrictions drastically reduced anthropogenic activities, resulting in a sharp decrease in emissions of air pollutants (Doumbia et al., 2020; Huang et al., 2020; Wang et al., 2020a).

The huge and large-scale emission reductions during the COVID-19 lockdown can be treated as a natural outdoor experiment to improve our understanding of the air pollutant's response to emission control. According to satellite and surface observations, compared with the period before the lockdown, nitrogen dioxide ($NO_2$) concentrations decreased by over 50% in China during the lockdown period(Bauwens et al., 2020; Liu et al., 2020; Shi and Brasseur, 2020; Zhang et al., 2020). The concentrations of other pollutants, including $SO_2$, particulate matter with an aerodynamic diameter less than 2.5 μm ($PM_{2.5}$), particulate matter with an aerodynamic diameter less than 10 μm ($PM_{10}$), and carbon monoxide (CO), also declined in a large area of China(Miyazaki et al., 2020; Wang et al., 2020b). However, surface ozone ($O_3$) concentrations in northern and central China increased by over 100%(Lian et al., 2020; Shi and Brasseur, 2020). Similar $O_3$ increases have been reported in southern Europe, India, and Brazil despite the large decrease in other pollutants(Sharma et al., 2020; Sicard et al., 2020; Siciliano et al., 2020). However, the underlying factors driving the $O_3$ changes during the city lockdowns remain unclear.

Surface $O_3$ is produced by photochemical reactions of ozone precursors, $NO_x$, volatile organic compounds (VOCs), and carbon monoxide (CO) and can also be transported from higher levels of the atmosphere and from outside regions (Akimoto et al., 2015; Liu and Wang, 2020a; Roelofs and Lelieveld, 1997). It is well known that $O_3$ has a non-linear dependence on its precursors and that $NO_x$ can either decrease or increase $O_3$ depending on the relative abundance of $NO_x$ to VOCs (Atkinson, 2000; Wang et al., 2017a). In general, the $O_3$ production in urban areas with high $NO_x$/VOCs ratios is VOCs limited, and reducing $NO_x$ emissions can increase $O_3$ due to decreased titration of $O_3$ and radicals. In addition to the two precursors, particulate matters can influence ozone by altering the solar irradiance and chemical reactions on aerosol surfaces (Li et al., 2019b; Liu and Wang, 2020b; Stadtler et al., 2018). Meteorological factors affect surface ozone by changing transport pattern, wet and dry depositions, chemical reaction rates, and natural emissions (Liu and Wang, 2020a; Lu et al., 2019).

The responses of ozone (and other air pollutants) to short-term emission reductions have been previously studied for a number of public and political events in China, such as the Beijing Summer Olympic Games (August 2008), the Asia-Pacific Economic Cooperation (APEC) meeting in Beijing (November 2014), and the G20 summit in Hangzhou (September 2016). During these events, various emission-reducing measures were implemented in the cities concerned and their surrounding areas. Whereas atmospheric concentrations of primary air pollutants ($NO_x$, CO, primary PM, and $SO_2$) in the concerned cities generally decreased in response to the temporary control measures, the $O_3$ concentrations showed mixed responses. $O_3$ decreased after emission reductions for some events (Huang et al., 2017; Wang et al., 2017b) but increased in others (Wang et al., 2010; Wang et al., 2015; Wu et al., 2019). The different $O_3$ responses have been qualitatively attributed to differences in the meteorological conditions (including regional transport of air masses) and to different control measures implemented by the local governments. Compared with the previously studied situations, the COVID-19 lockdown is unique in that emissions decreased across the whole country (and later worldwide) as opposed to a specific city or region, and the decreases were also much more drastic than those due to transportation restrictions alone. Moreover, the COVID-19 lockdown took place in winter, whereas the previous interventions occurred in summer and autumn, when meteorology and atmospheric chemistry are different from winter. The present study analyzes surface $O_3$ data across China before and during the COVID-19 lockdown. We find that $O_3$ decreased in southern China while increasing in most other regions during the lockdown. Using a regional chemistry transport model, we isolate the impacts of meteorological changes and anthropogenic emission reductions on $O_3$. Our results highlight the importance of meteorological influences on the short-term $O_3$ changes and the diverse response of $O_3$ to the emission reductions of its precursors in different climate and emission-mix regions.

## 2 Materials and Methods

### 2.1 Surface measurement data

We obtained the observed concentrations of surface $O_3$ and other pollutants ($PM_{2.5}$, $PM_{10}$, $SO_2$, CO, $NO_2$) at 1643 stations from the China National Environmental Monitoring Center (http://106.37.208.233:20035/). Data quality control was conducted for the measurement data in accordance with previous studies(Lu et al., 2018; Song et al., 2017). Fig. 1 shows the locations of these environmental monitoring sites.

The country-wide measures to control the spread of COVID-19 were implemented starting from 23 January 2020 (the exact date varies for different cities), just before the Chinese New Year. All enterprises remained closed until no earlier than 13 February, except those required for essential public services, epidemic prevention and control, and residential life needs. We focused on the period during the COVID-19 lockdown from 23 January to 12 February 2020 (hereafter referred to as the CLD period), 3 weeks in total. We derived the changes in $O_3$ and other pollutants by comparing the CLD period with the 3 weeks before the COVID-19 outbreak, from 2 to 22 January 2020 (hereafter referred to as the pre-CLD period). We focused on three typical regions in China (Fig. 1): north China (NC, 35-41.5°N, 113-119°E, including Beijing, Tianjin, Hebei, and western Shandong), central China (CC, 28.8-33°N, 108-117°E, including Hubei province, where Wuhan is situated, and the surrounding regions), and south China (SC, 21.5-24°N, 111-116°E, including the Pearl River Delta and the surrounding regions). The NC region is situated in the North China Plain, which is known to suffer from severe haze in winter; CC was the original epicenter of the COVID-19 outbreak in China and is an important economic hub for the central regions of China; the Pearl River Delta, where the megacities of Guangzhou and Shenzhen are situated, is the most developed region in southern China.

**2.2 Numerical modeling**

The CMAQ model (Community Multiscale Air Quality model, v5.2.1) was applied to simulate the $O_3$ mixing ratios over China from 2 January to 12 February 2020. The WRF model (Weather Research and Forecasting model, v3.5.1) was driven by the dataset of the National Center for Environmental Prediction (NCEP) FNL Operational Model Global Tropospheric Analyses with a horizontal resolution of 1° × 1° and provided meteorological inputs for the CMAQ model. The CMAQ target domain covered the continental China at a horizontal resolution of 36 km × 36 km. SAPRC07TIC (Carter, 2010; Hutzell et al., 2012) and AERO6i (Murphy et al., 2017; Pye et al., 2017) were adopted as the gas-phase chemical mechanism and aerosol mechanism, respectively. The CMAQ model has been improved with updated heterogeneous reactions to better predict the $O_3$ concentration; details can be found in Liu and Wang (2020a). Although the WRF-CMAQ model was run in offline mode, the CMAQ model employs an in-line method that uses the concentrations of particles and $O_3$ predicted within a simulation to calculate the solar radiation and photolysis rates (Binkowski et al., 2007). As a result, the effect of aerosol on $O_3$ concentrations via changing the photolysis rates were also considered in the simulation. The chemical boundary conditions were provided by the results of Whole Atmosphere Community Climate Model (WACCM, https://www.acom.ucar.edu/waccm/download). The anthropogenic emissions in China were obtained from the Multi-resolution Emission Inventory for China (MEIC) in 2017 (http://www.meicmodel.org) with scaling factors to the year 2020 (Table S1, see text in Supplementary Information), which were estimated based on the Three-Year Action Plan (2018–2020) issued by the government and the changes in the multi-pollutant emissions of different sectors in recent years (Zheng et al., 2018). The emission adjustments during the lockdown period are based on recent publications (see text in Supplementary Information). Emissions from the other countries were derived from the MIX emission inventory (Li et al., 2017). International shipping emissions were taken from the Hemispheric Transport Atmospheric Pollution (HTAP) emission version 2.2 dataset for 2010(Janssens-Maenhout et al., 2015). Biogenic emissions were calculated by the Model of Emissions of Gas and Aerosols from Nature (MEGAN) version 2.1(Guenther et al., 2012) with meteorological inputs from the WRF model.

Two experiments were conducted to investigate the impacts of meteorological changes and emission reductions on $O_3$ during the CLD period. The first (baseline) used the same anthropogenic emissions for the pre-CLD and CLD periods, and the second (Reduction case) used emission reductions of 70%, 40% and 30% for transportation, industry and power generation, respectively, and a 10% increase of residential emission during the CLD period. These emission reductions for the whole country were estimated according to the previous literature(Doumbia et al., 2020; Huang et al., 2020; Wang et al., 2020a). Comparing these two model simulations, the $O_3$ changes during the CLD period relative to the pre-CLD period for the Reduction Case were considered to be entirely due to the meteorological changes and emission reductions. The impacts of the meteorological changes (including the changes in chemical boundary conditions) were quantified by subtracting the $O_3$ mixing ratios of the pre-CLD period from those of the CLD period for the baseline experiment, while the impacts of emission reduction were estimated by comparing the $O_3$ mixing ratio during the CLD period between the Reduction Case and the baseline experiment. Furthermore, we individually reduced the emissions of nitrogen oxide ($NO_x$), VOCs, CO, PM (particulate matter, including $PM_{10}$, $PM_{2.5}$, black carbon, and organic carbon), and $SO_2$ during the CLD period to elucidate the response of $O_3$ to each pollutant reduction.

The performance of the CMAQ model in simulating the $O_3$, $NO_2$, $PM_{2.5}$, $SO_2$, and CO concentrations for the Reduction Case was evaluated (Fig. S1 and Table S2), showing reasonable agreements with the respective surface observations. Details of the emission estimation and the model evaluation are presented in Supplementary Information.

## 3 Results

### 3.1 Observed $O_3$ changes in different parts of China

Figures 2 and 3 present the changes in observed concentrations of $O_3$ and other pollutants during the CLD period compared with pre-CLD. The concentrations of most pollutants ($SO_2$, CO, $PM_{2.5}$, $PM_{10}$) that partially or fully originated from the direct emissions declined in China during the lockdown. $NO_2$, a precursor of $O_3$, decreased by about 50% across the entire continental China, and by similar amounts in all regions (Fig. 3b). However, the $O_3$ mixing ratio exhibited varying changes in different regions (Fig. 2a). In NC and CC, the daily average $O_3$ increased significantly, by 112% and 73%, respectively (Fig. 2b); in contrast, it remained almost unchanged in SC. The $O_3$ changes also varied between daytime (8:00-20:00 LST) and nighttime (20:00-8:00 LST). During daytime, the $O_3$ increase in most parts of China was smaller than the daily average (Fig. 2c), 92% and 71% in the NC and CC regions (Fig. 2d). In the SC region, most stations displayed a decrease in $O_3$ during daytime, leading to a regional average $O_3$ drop of 14%. During nighttime, the $O_3$ mixing ratio increased significantly across China (Fig. 2e), by 154%, 77%, and 18% in NC, CC, and SC, respectively (Fig. 2f). These results reveal the diverse response of $O_3$ during the lockdown in different regions, especially for the daytime. The changes in the $O_x$ ($NO_2+O_3$) concentration (Fig. 3a), which takes into account the NO titration, also varied in different regions. The daytime average $O_x$ levels increased by 4% in NC and by 11% in CC, and decreased by 29% in SC. These results suggest that the NO titration effect was not the only cause of the $O_3$ increase in northern and central China, as $O_x$ would have decreased with sharply reduced $NO_x$ emissions.

**3.2 Contribution of meteorological changes and emission reductions to $O_3$**

Ground-level $O_3$ is influenced by both chemical reactions of $O_3$ precursors and meteorology. In this study, we used the WRF-CMAQ model to separate the impacts of meteorological changes and emission reductions on the changes in $O_3$ across China (Fig. S2), which reveals significant contributions of both meteoroglgy (over most of continental China) and emissions (mainly in populated areas of eastern China) . Fig. 4 shows the more detailed results for the NC, CC, and SC regions for both daytime and nighttime and Fig. S3 presents the meteorological impact and emission impact in terms of percent change. The observed $O_3$ changes in these regions were reasonably captured by the simulations. For the daytime average, the $O_3$ increase in NC was attributed to the comparable contributions from both meteorological changes (58%) and emission reductions (42%) (Fig. 4a). In CC (Fig. 4b), the meteorological change (98%) was the primary cause of the $O_3$ increase, whereas the contribution of emission reduction was much lower (2%). In SC (Fig. 4c), the meteorological changes (73%) and emission reductions (27%) both contributed to the $O_3$ decrease. During nighttime, the emission reduction increased $O_3$ in all three regions (including SC), and its impact was stronger; the effect of meteorological changes weakened at night (Fig. S3).

**3.3 Impacts of meteorological changes on $O_3$**

The impacts of meteorological changes on $O_3$ for the NC, CC, and SC regions can be explained by the changes in the weather pattern and specific meteorological factors. In winter, continental China is generally controlled by a cold high-pressure system (Fig. 5). During our study period, the center of this high-pressure system was located in northern China, moving southward from the pre-CLD to the CLD period, with weakening strength. The high-pressure system therefore became increasingly dominant in southern China, and the strengthened southward winds brought colder air masses from the north (Fig. 6c), which decreased the temperature locally (Fig. 6a). In contrast, in central and northern China, the winds shifted to a more northward direction, transporting warmer air masses from the south (Fig. 6c), which increased the temperature (Fig. 6a). During daytime, the decrease (increase) in temperature in the SC region (CC and NC regions) weakened (enhanced) the surface $O_3$ chemical production. Biogenic emission is an important source of VOCs and thereby contributes to $O_3$ formation in China (Wu et al., 2020). The temperature changes led to an increase (decrease) of biogenic emissions in the CC (SC) region (Fig. S4). Thus, the temperature changes increased (decreased) $O_3$ in the CC (SC) region by influencing chemical reaction rates directly (Fu et al., 2015; Steiner et al., 2010) and altering biogenic emissions indirectly (Im et al., 2011; Liu and Wang, 2020a).

The changes in the weather pattern also resulted in less clouds and precipitation in northern and central China, but more clouds and precipitation in southern China (Fig. 6e and f). Clouds can reduce the amount of solar radiation reaching the surface and thus the chemical production of $O_3$(Lelieveld and Crutzen, 1990), while precipitation can also reduce $O_3$ through the scavenging of its precursors(Seinfeld and Pandis, 2006; Shan et al., 2008). The cloud and precipitation patterns therefore contributed to $O_3$ increases in CC and NC and decreases in SC. Furthermore, in NC and CC, the significant increase in the planetary boundary layer height during the lockdown (Fig. 6d) might promote the transport of $O_3$ from the upper air to the surface, contributing to the $O_3$ increase in these regions (He et al., 2017; Sun et al., 2009). The increase (decrease) in specific humidity in NC and CC (SC) might also have contributed to the decrease (increase) in $O_3$ mixing ratios in those regions (Li et al., 2019c; Ma et al., 2019) (Fig. 6b). During nighttime, the changes in meteorological factors were similar to those in daytime (Fig. S5) but exerted smaller impacts on $O_3$ changes due to the decreasing effects of temperature and cloud cover (negligible biogenic emissions and solar radiation).

**3.4 Response of $O_3$ to emission reductions**

We further investigated the impact of multi-pollutant reductions on the $O_3$ changes. Because transportation and industrial activities were reduced significantly during the lockdown and they are the major sources of $NO_x$ (>80%) and VOCs (>60%) (Fig. 7), the estimated reductions of $NO_x$ and VOC emissions were more significant than those for CO, particulate matter (PM), and $SO_2$ (Fig. 8a, c, e). The $NO_x$ emission reductions were 46%, 49%, and 55% in the NC, CC, and SC regions, respectively, while the respective reductions for VOC emissions were 32%, 24%, and 45%. The relationship between $O_3$ and the emissions of its precursors is non-linear. We used the ratio of production rates between $H_2O_2$ and $HNO_3$ ($P_{H2O2}/P_{HNO3}$) (Gaubert et al., 2021; Tonnesen and Dennis, 2000) to identify the $O_3$ formation regime in China for the periods before and during the lockdown (Fig. 9). $P_{H2O2}/P_{HNO3}<0.06$ is VOC-limited region; $P_{H2O2}/P_{HNO3}\geq0.2$ is $NO_x$-limited region, and $0.06\leq P_{H2O2}/P_{HNO3}<0.2$ is the transition zone. For the pre-CLD period, during daytime, the VOC-limited (or $NO_x$-saturated) regions included North China Plain and other urban areas, while $NO_x$-limited regions located in southern China and other rural areas (Fig. 9a). During nighttime, most regions are VOC-limited (Fig. 9c).

The $O_3$ formation regime determines the response of $O_3$ to the $NO_x$ reduction during the CLD period. During daytime, $NO_x$ reduction increased $O_3$ in $NO_x$-saturated regions, but decreased it in $NO_x$-limited regions (Fig. 10b). We also found that although the daytime $O_3$ formation regime in most regions shifted from the VOC-limited regime to the $NO_x$-limited regime after the emission reductions during the CLD period, the daytime $O_3$ formation in the North China Plain was still controlled by the VOC level (Fig. 9b), which suggests that the $NO_x$ level is still high in this region. During nighttime, the reduction of $NO_x$ emission contributed increased $O_3$ due to the NO titration effect in a large areas (Fig. 10c). The reduction of VOC emission decreased $O_3$ across China (Fig. 10d-f). As an $O_3$ precursor, the reduction of CO emission also contributed to a small decrease in the $O_3$ mixing ratio (Fig. 10g-i); in contrast, the PM and $SO_2$ emissions reductions increased $O_3$ (Fig. 10j-o) through the weakening of aerosol effects (Li et al., 2019a; Liu and Wang, 2020b), but their impacts were much smaller and were insignificant (< 1 ppbv) due to the smaller reductions, compared with $NO_x$ and VOCs.

The response of $O_3$ to the emission reductions in different regions depended on the levels of $NO_x$ and VOC reductions. For the daytime average, in the $NO_x$-saturated NC region, the $O_3$ increase by the $NO_x$ reduction counteracted the $O_3$ decrease by the VOC emission reduction, leading to the decrease in increased $O_3$ production rates (Fig. 11) and a substantial net $O_3$ increase (Fig. 8b). In CC, the contributions of the $NO_x$ and VOC reductions were comparable in magnitude, and their opposing impacts resulted in only a slight change in $O_3$ (Fig. 8d). In the $NO_x$-limited SC region, the impact of the $NO_x$ reduction on $O_3$ was smaller than that of the reduction of VOCs, leading to the decrease in $O_3$ production rates (Fig. 11), and a net decrease in $O_3$ (Fig. 8f). During nighttime, the effect of the VOC reduction was weakened due to the lower rate of degradation of VOCs by radicals compared with daytime, and the $O_3$ level increased in all three regions due to decreases in the NO titration effect (Fig. 11). The impacts of emission reductions on whole-day average $O_3$ were similar to those during daytime.

The above modeling results show that the contribution of $NO_x$ reductions (by 46%–55%) to the rise of $O_3$ decreased from NC

to CC and to SC, reflecting the decreasing level of $NO_x$ saturation from north to south. In contrast, the impact of the estimated VOC reduction on $O_3$ increased from north to south, which can in part be attributed to the regional variation of VOC reductions. In the SC region, transportation and industry are the predominant sources of VOCs (97%, compared with 85% and 60% in NC and CC, respectively) (Fig. 7). During the CLD period, the reduction of VOC emission in SC (45%) was significant and comparable with the $NO_x$ reduction (55%). In contrast, the VOC reductions in the NC (32%) and CC (24%) regions were much lower (Fig. 8a, c) and could not offset the impact of $NO_x$ reduction on $O_3$. The residential sector (mainly household coal burning) is an important source of VOC emission in the NC and CC regions, whereas its contribution is smaller in SC. The residential emissions increased during the CLD period because many migrant workers came back for the Chinese New Year holiday and were stranded there due to the lockdown (Wang et al., 2020a; Wang et al., 2020b).

## 4. Conclusion

The first country-wide lockdown during the COVID-19 outbreak in China drastically reduced transportation and industrial activities, leading to sharp declines in air pollutant emissions from these sectors. Surface $O_3$ in urban areas of China responded differently in the northern (increase) and southern regions (decrease) compared to the three-week period before the lockdown, which can be explained by changes in meteorology and differences in the $O_3$ chemistry regimes and the magnitudes of precursor reductions in these regions. The model simulated contributions of meteorology to daytime $O_3$ changes were larger or comparable to most regions. The extent of VOC reduction, which suppressed $O_3$ formation, was insufficient to offset the large NO titration effect during daytime in northern China, and that larger reductions of VOCs (e.g., from residential sectors) would have been needed to reduce the $O_3$ concentration in the northern and central China. The rising $O_3$ concentration in northern China during the COVID-19 lockdown and in recent winters should receive greater attention because $O_3$ boosts the atmospheric oxidative capacity and therefore production of secondary aerosols (Fu et al., 2020; Huang et al., 2020; Zhu et al., 2020), which are important components of winter haze in northern China. Our findings in China are relevant to untangling the underlying factors driving the $O_3$ changes in other parts of the world during their COVID-19 lockdowns.

**Data availability**

The codes and data used in this study are available upon request from Yiming Liu (liuym88@mail.sysu.edu.cn) and Tao Wang (cetwang@polyu.edu.hk).

**Acknowledgements**

This work was supported by the Hong Kong Research Grants Council (T24-504/17-N and A-PolyU502/16) and the National Natural Science Foundation of China (91844301). B.G. acknowledges support by the National Center for Atmospheric Research, which is a major facility sponsored by the National Science Foundation under cooperative agreement no. 1852977. We would like to thank Prof. Qiang Zhang from Tsinghua University for providing the emission inventory.

**Author contributions**

T.W. initiated the research. Y.M.L. and T.W. designed the research framework. C.G., and T.D. estimated the emission changes. Y.M.L. performed model simulations and drew the figures. T.W. and Y.M.L. analyzed the results. T.W. and Y.M.L wrote the paper with the contributions from all the authors.

**Competing interests**

The authors declare that they have no conflict of interest.

**Materials & Correspondence**

Correspondence and requests for materials should be addressed to T.W. or Y.M.L.

**Additional information**

Supplementary information is available for this paper.

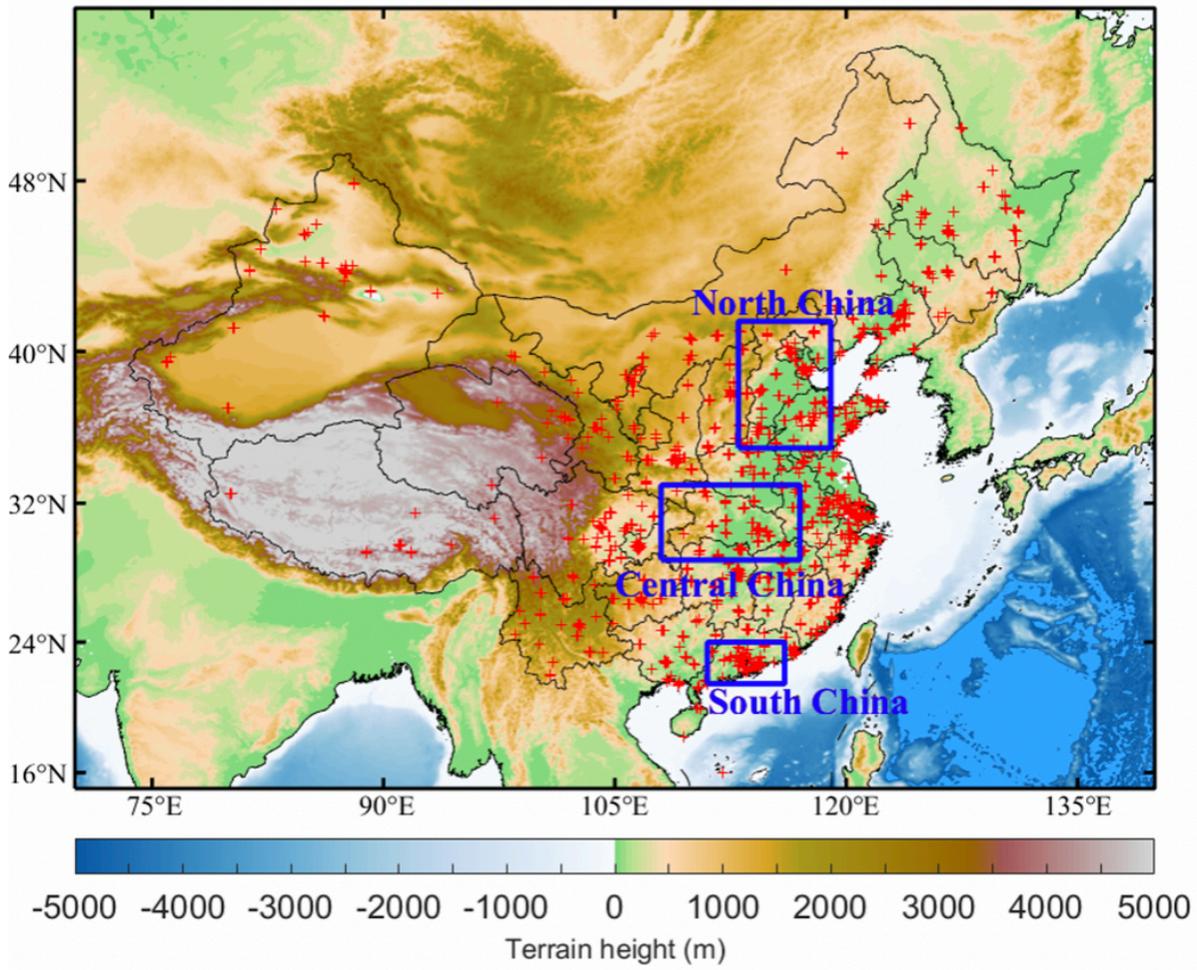

Figure 1: Location of 1643 environmental monitoring stations (red "+" symbols) operated by the Ministry of Ecology and Environmental Protection of China. The blue boxes denote the regions of north China, central China, and south China designated for further analysis.

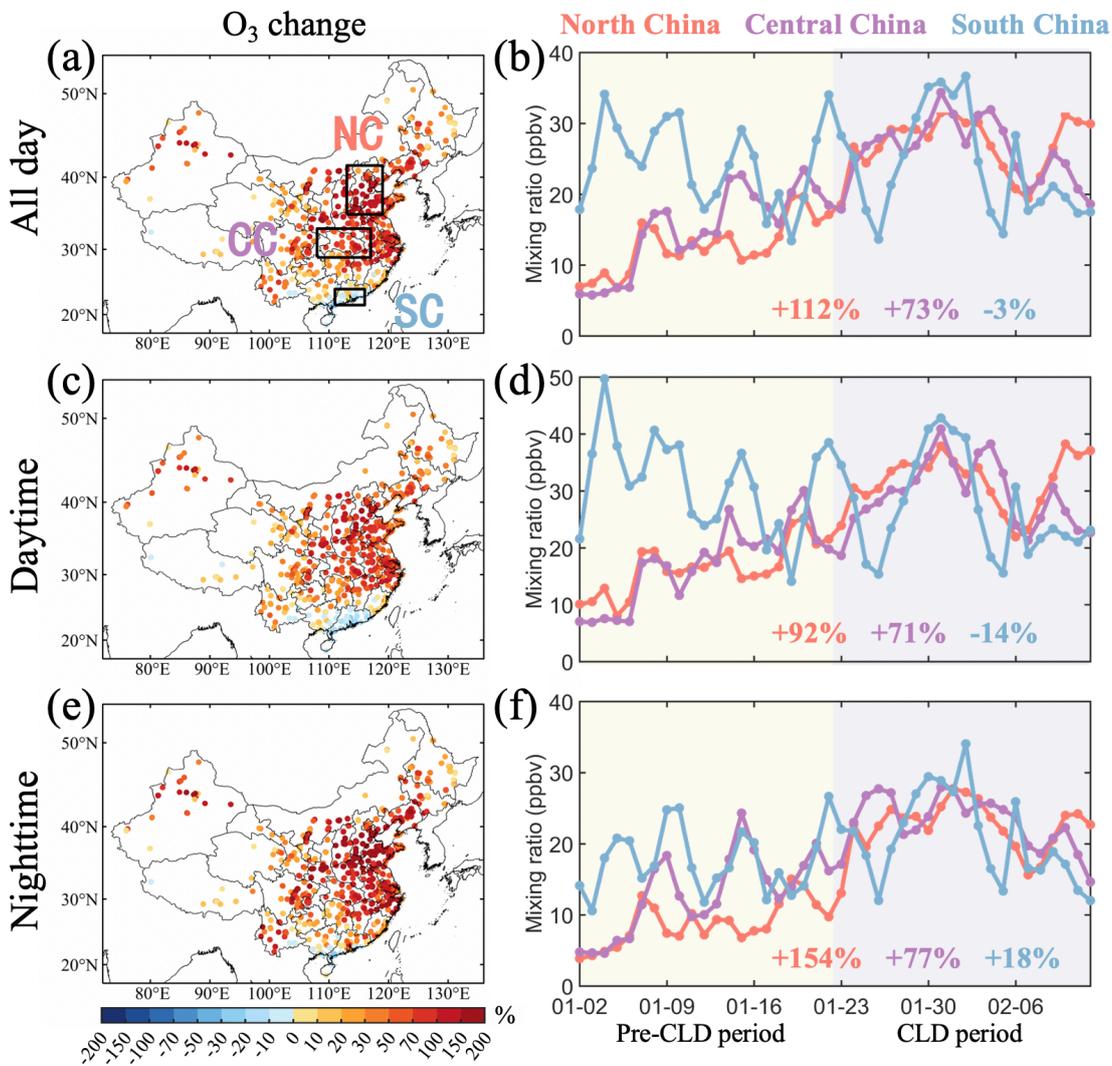

Figure 2: Observed changes in $O_3$ mixing ratios across mainland China before and during the COVID-19 lockdown period. (a, c, e) The spatial distribution of $O_3$ changes for all-day average, daytime average, and nighttime average during the CLD period compared with the pre-CLD period. The black boxes in (**a**) show the locations of north China (NC, 184 sites), central China (CC, 108 sites), and south China (SC, 77 sites). (**b**) The variations of all-day average $O_3$ mixing ratios during the study period for the NC, CC, and SC regions. (**d**) The same with (**b**) but for daytime average $O_3$. (**f**) The same with (**b**) but for nighttime average $O_3$.

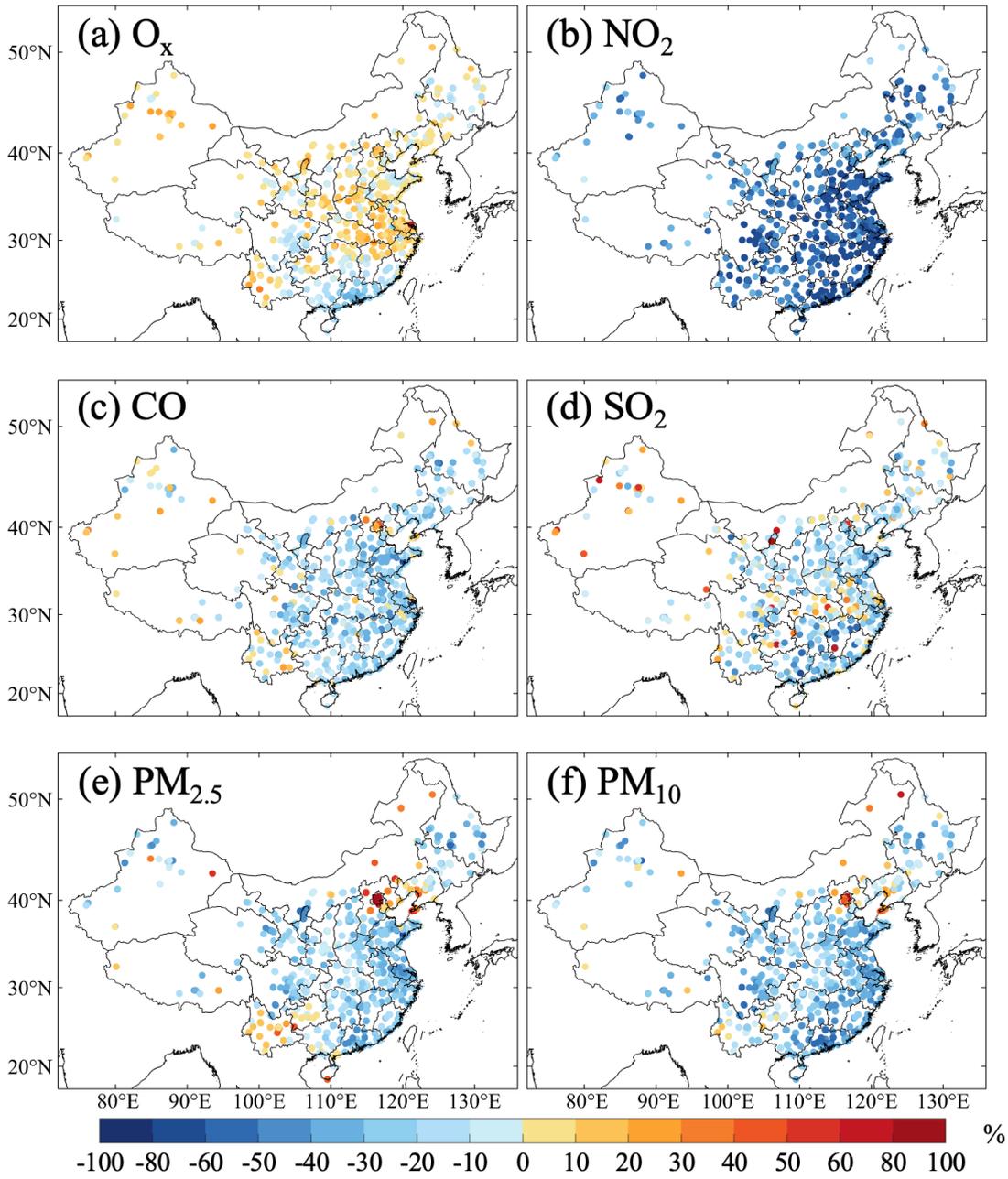

Figure 3: Percentage change of (a) observed daytime average $O_x$ ($NO_2+O_3$), whole-day average (b) $NO_2$, (c) CO, (d) $SO_2$, (e) $PM_{2.5}$, and (f) $PM_{10}$ concentrations during the CLD period relative to the pre-CLD period.

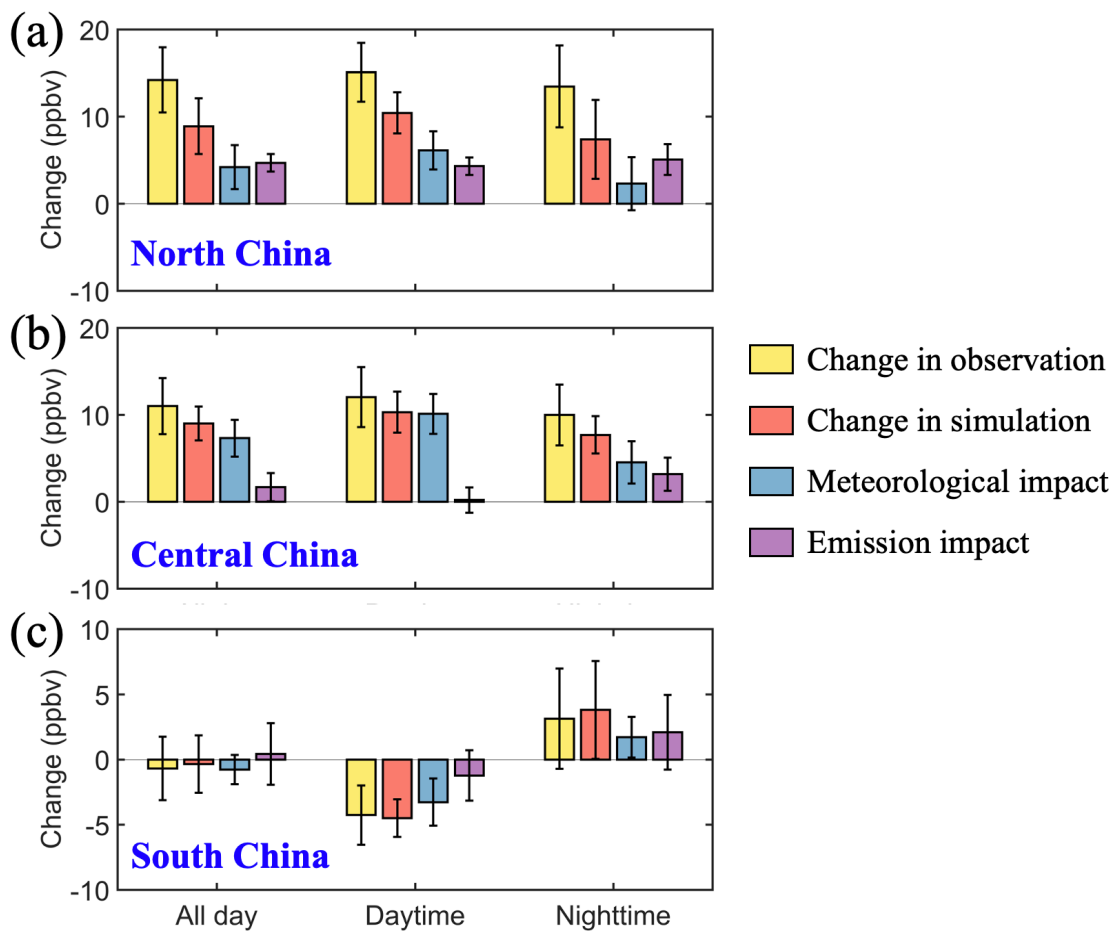

Figure 4: Changes in $O_3$ mixing ratios during the COVID-19 lockdown period and contributions from meteorological changes and emission reductions for three typical regions. (a) Observed and simulated changes in $O_3$ mixing ratios and the contributions from meteorological changes and emission reductions during the CLD period compared with the pre-CLD period in north China (NC). The $O_3$ changes for the all-day average, daytime average, and nighttime average are presented. (b) The same with (a) but for central China (CC). (c) The same with (a) but for south China (SC). The locations of these three regions are shown in Fig. 1. Note that the error bars mark the standard deviations within the region.

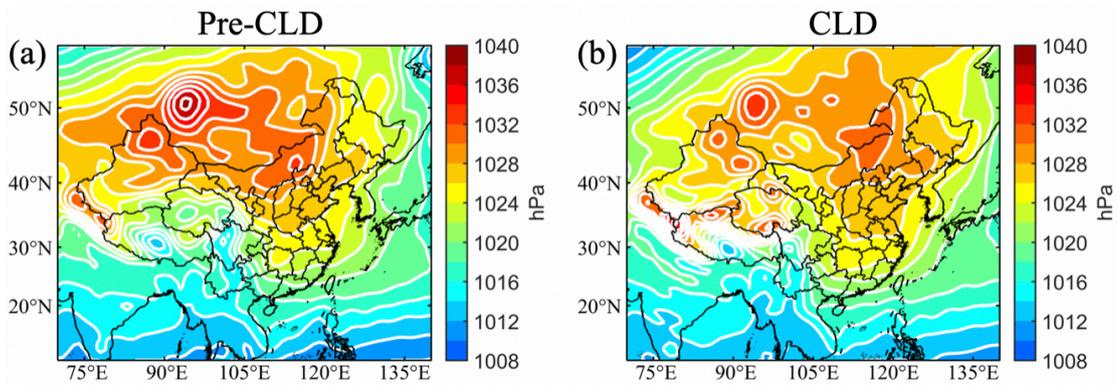

Figure 5: Averaged sea-level pressure during the pre-CLD and CLD periods. Data are from the National Center for Environmental Prediction (NCEP) FNL Operational Model Global Tropospheric Analyses dataset.

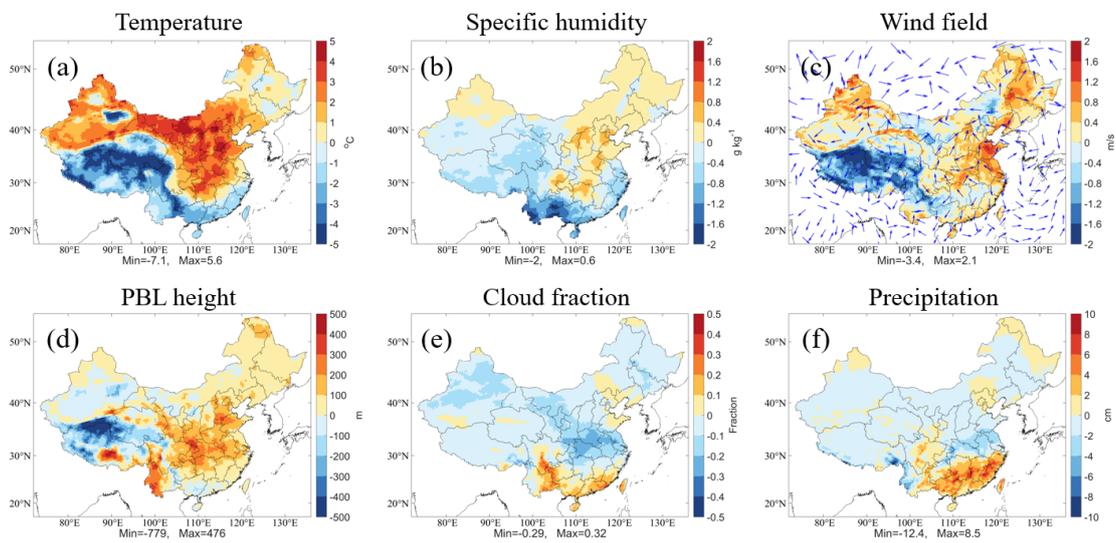

Figure 6: Model simulated changes in daytime temperature at 2 m height, specific humidity at 2 m height, wind field at 10 m height, planetary boundary layer (PBL) height, cloud fraction, and precipitation during CLD period relative to pre-CLD period. In panel (c), the shaded color and vector represent the wind speed and wind direction, respectively.

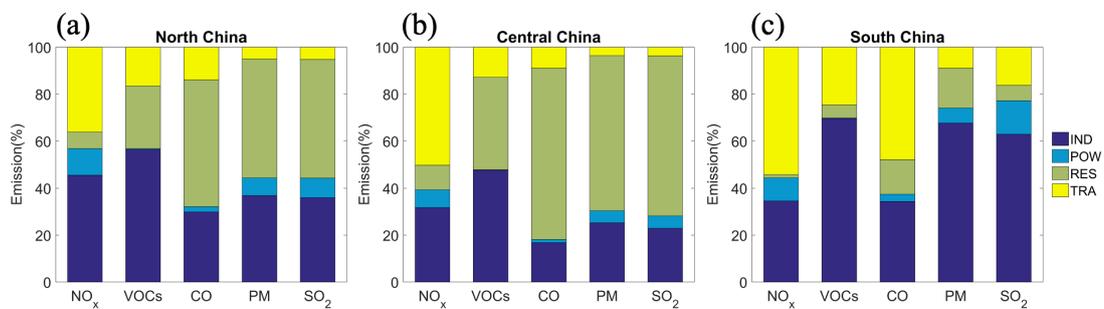

Figure 7: Percentage contribution to $NO_x$, VOCs, CO, PM, and $SO_2$ emissions from industrial (IND), power plant (POW), residential (RES), and transportation (TRA) sectors in (a) north China, (b) central China, and (c) south China. Emission data are from 2017 MEIC (http://meicmodel.org) with estimated scaling factors from 2017 to 2020.

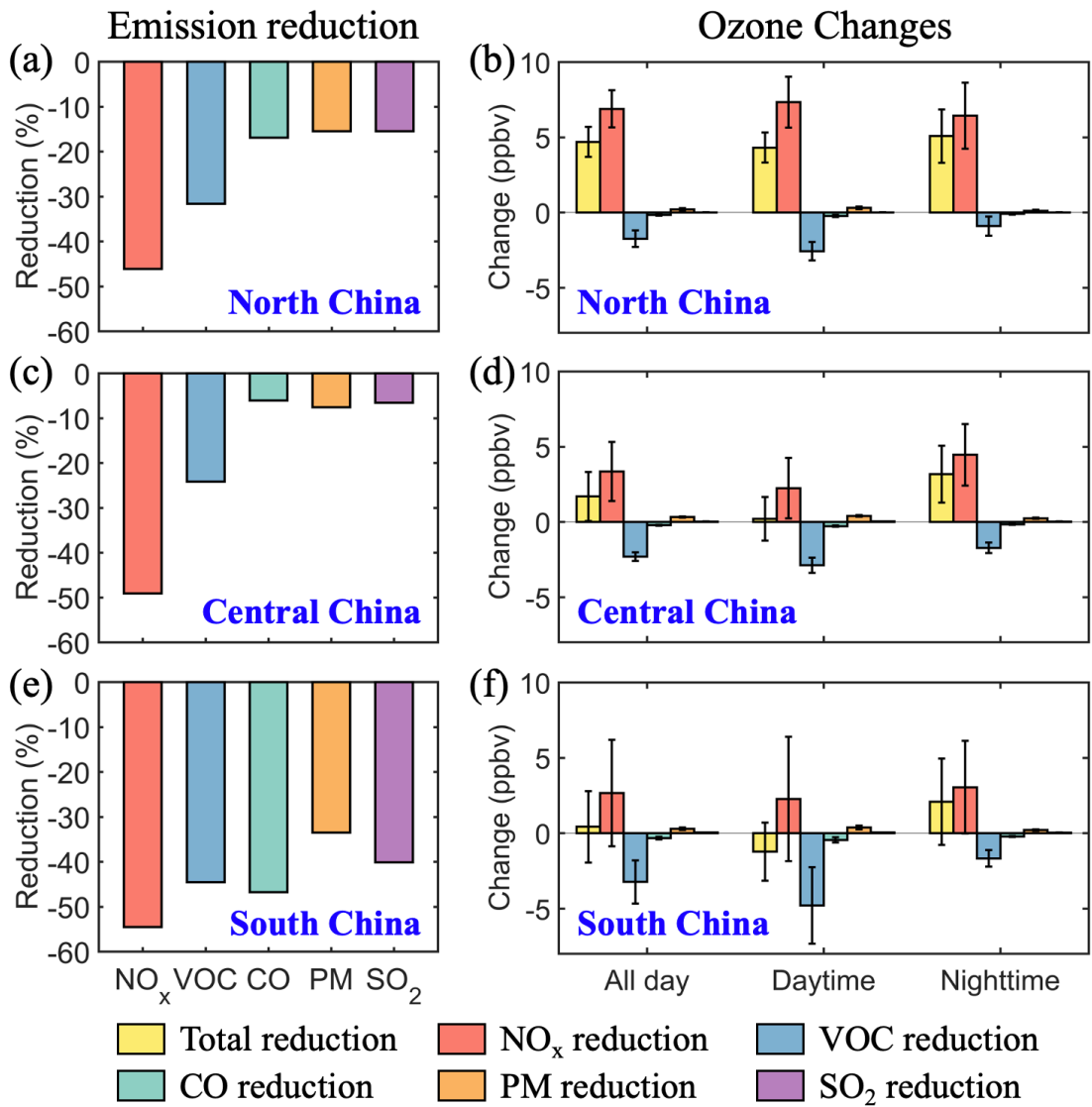

Figure 8: The estimated reductions of multi-pollutant emissions due to the COVID-19 lockdown and their impacts on the $O_3$ changes for three regions. (a, c, e) The estimated reductions of $NO_x$, VOC, CO, PM, and $SO_2$ emissions during the CLD period compared with the pre-CLD period for north China, central China, and south China. (b, d, f) The impacts of different pollutant emission reductions due to the lockdown on $O_3$ changes for the three regions. The $O_3$ changes for all-day average, daytime average, and nighttime average are presented. The error bars are the standard deviations.

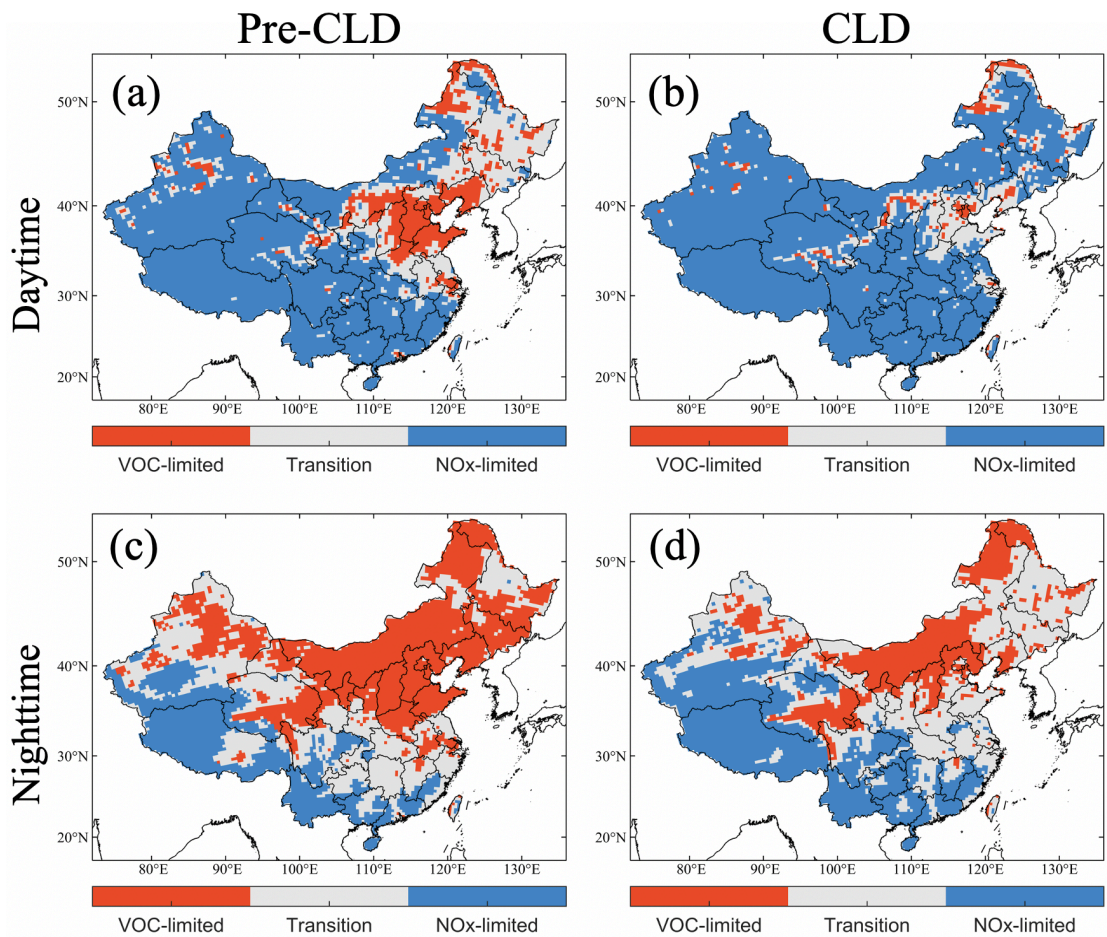

Figure 9: Ozone formation regime in the daytime and nighttime before and during the lockdown periods estimated by the ratio of the production rates of hydrogen peroxide to nitric acid ($P_{H2O2}/P_{HNO3}$). VOC-limited region: $P_{H2O2}/P_{HNO3}<0.06$; NO$_x$-limited region: $P_{H2O2}/P_{HNO3} \geqslant 0.2$, Transition zone: $0.06 \leqslant P_{H2O2}/P_{HNO3}<0.2$. The production rates of $H_2O_2$ and $HNO_3$ were calculated by the integrated reaction rate (IRR) diagnose tool in the CMAQ model.

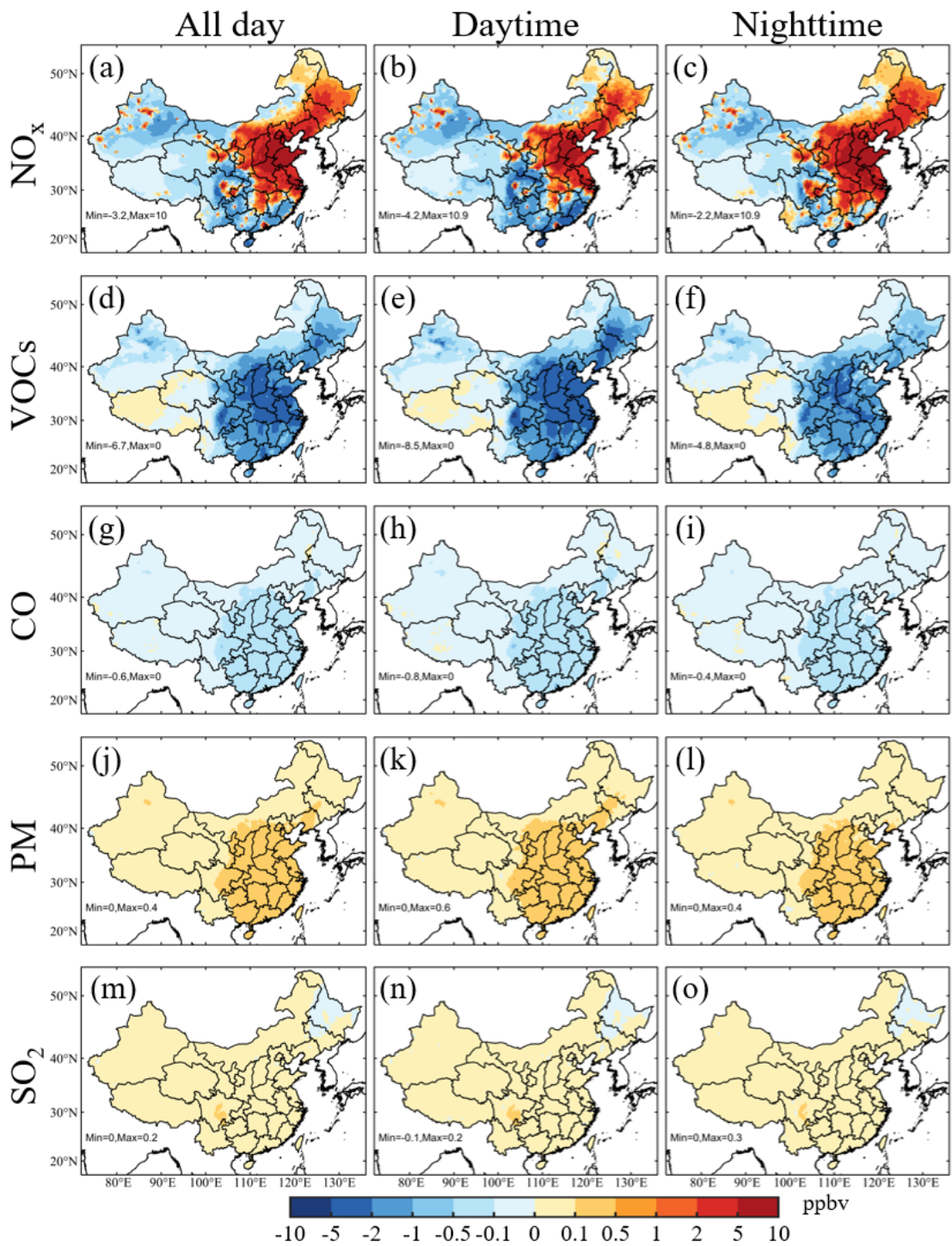

Figure 10: Model simulated changes in O$_3$ mixing ratios for all-day average, daytime average, and nighttime average due to the reductions of NO$_x$, VOC, CO, PM, and SO$_2$ emissions during the CLD period compared with the pre-CLD period.

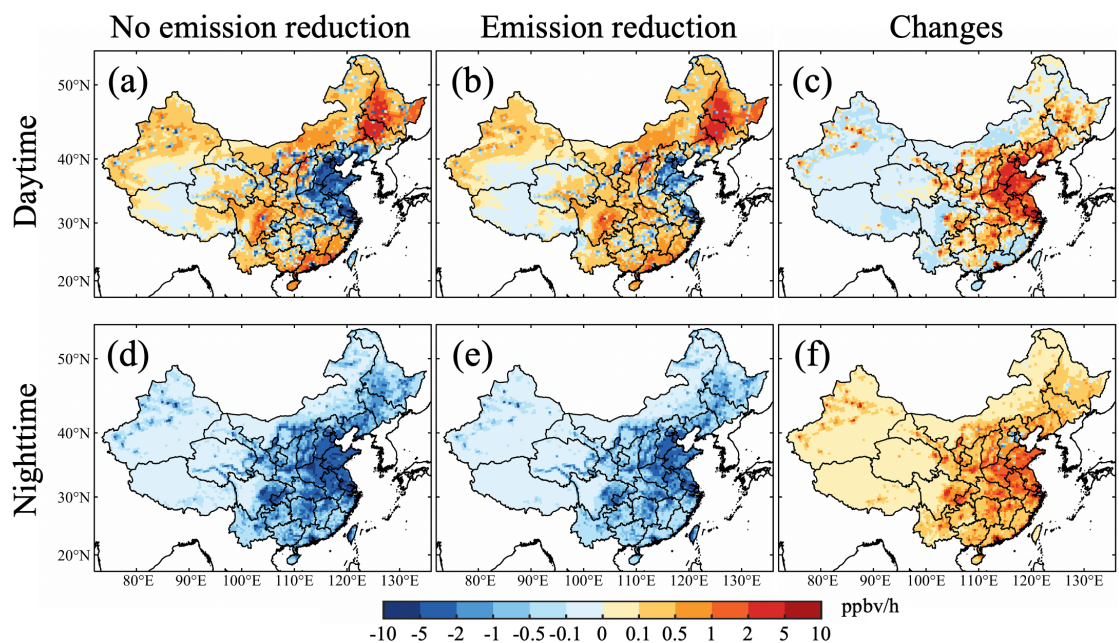

Figure 11: O$_3$ chemical production rates before and after the anthropogenic emission reductions and the changes during the COVID-19 lockdown period. The chemical production rates were calculated by the process analysis method in the CMAQ model.

Supplementary Information for

"Diverse response of surface ozone to COVID-19 lockdown in China"

Yiming Liu[a,*#], Tao Wang[a,*], Trissevgeni Stavrakou[b], Nellie Elguindi[c], Thierno Doumbia[c], Claire Granier[c,d], Idir Bouarar[e], Benjamin Gaubert[f], Guy P. Brasseur[a,e,f]

[a]Department of Civil and Environmental Engineering, The Hong Kong Polytechnic University, Hong Kong, China;

[b]Royal Belgian Institute for Space Aeronomy, Brussels, Belgium;

[c]Laboratoire d'Aérologie, Toulouse, France;

[d]NOAA Chemical Sciences Laboratory and CIRES/University of Colorado, Boulder, CO, USA;

[e]Environmental Modeling Group, Max Planck Institute for Meteorology, Hamburg, Germany;

[f]Atmospheric Chemistry Observations and Modeling Laboratory, National Center for Atmospheric Research, Boulder, CO, USA;

*Correspondence to: Tao Wang (cetwang@polyu.edu.hk) and Yiming Liu (liuym88@mail.sysu.edu.cn)

#Now in School of Atmospheric Sciences, Sun Yat-sen University, Zhuhai, China

This file includes Supplementary text, and Supplementary Figures S1-S5, Supplementary Table S1-S2.

# Supplementary text
## Estimation of anthropogenic emissions in 2020

We estimated the anthropogenic emission for China in 2020 based on the MEIC 2017 (http://www.meicmodel.org) according to the control plan established by the Chinese government and the emission trends in recents years. From 2013 to 2017, the Chinese government launched the to Air Pollution Prevention and Control Action Plan to mitigate haze events. Zheng et al. (2018) compiled the trends of anthropogenic emissions during this period and demonstrated that the $SO_2$, $NO_x$, and PM (particulate matter, including $PM_{10}$, $PM_{2.5}$, and its components) emissions have been reduced significantly. In 2018, the Chinese government issued a Three-Year Action Plan (2018–2020) to further reduce the $SO_2$, $NO_x$, and PM emissions (http://www.gov.cn/zhengce/content/2018-07/03/content_5303158.htm). Previous control measures were implemented and the emissions were thought to continue decreasing after 2017. As the most recent available data on China's anthropogenic emissions are from 2017, we estimated the emissions for 2020 based on the MEIC 2017 according to the trends of these emissions in recent years (Zheng et al., 2018). Table S1 shows the scaling factors from 2017 to 2020 for $NO_2$, PM, and $SO_2$ in the power plant, industry, transportation, and residential sectors. The VOC emission was assumed to be unchanged from 2017 to 2020 because it increased by only 2% from 2013 to 2017 (Zheng et al., 2018). The $NO_2$, PM, and $SO_2$ emissions from transportation were also assumed constant from 2017 to 2020, considering that they had changed little during 2015–2017. The $NO_x$ emission in the residential and industrial sectors was assumed to be the same in 2020 as in 2017 in view of its flat trend in recent years. Because the $NO_x$ emission from power plants decreased by ~47% from 2013 to 2017 (11.7% per year), we assumed it to further decrease by 35% from 2017 to 2020. The same approaches were applied to the reductions of the PM and $SO_2$ emissions from 2017 to 2020. With these adjustments, we derived an estimated anthropogenic emission inventory for China in 2020. The model-simulated pollutant concentrations using this inventory showed a reasonable agreement with the surface measurement data during the period before the COVID-19 lockdown (Table S2, also see the Model evaluation section below), which suggested the estimated emission inventory was reasonable.

## Estimated reduction of anthropogenic emissions during the CLD period

We estimated the emission reductions during the COVID-19 lockdown period according to the recent literature (Doumbia et al., 2021; Wang et al., 2020; Huang et al., 2020). For the transportation sector, the decrease in national traffic volume was estimated at 70% during the lockdown according to transportation index data. The industry emissions were assumed to decrease by 40% across China. The emissions of power plants were estimated to decrease by 30% due to the less consumption of electricity. The emissions of residential activities were assumed to increase by 10% due to more coal combustion and cooking. We kept the emissions from agriculture unchanged because they were less affected by the city lockdowns. Fig. 8 presents the reductions of $NO_x$, VOCs, CO, PM, and $SO_2$ emissions due to the changes in anthropogenic activities during the lockdown, which generally match the estimation by Huang et al. (2020). The observed pollutant concentrations during the lockdown period could faithfully captured by the CMAQ model using this estimated reduced emissions (Table S2, also see the Model evaluation section below), which suggested the estimated reductions of anthropogenic emissions were acceptable.

## Model evaluation

Statistical parameters were calculated to validate the model performance in simulating the air pollutant concentrations from January 2 to February 12, 2020, including the mean observation (OBS), mean simulation (SIM), mean bias (MB), mean absolute gross error (MAGE), root mean square error (RMSE), index of agreement (IOA), and correlation coefficient ($r$). The equations of these statistical parameters can be found in Fan et al. (2013).

Table S2 shows the evaluation results of the simulated concentrations of $SO_2$, $NO_2$, CO, $O_3$ and $PM_{2.5}$ in China for the periods before and during the COVID-19 lockdown, respectively. As the measured $NO_2$ by the catalytic conversion method in the national network overestimates the $NO_2$ (Xu et al., 2013), we adjusted the observed $NO_2$ data following the method proposed by Zhang et al. (2017) and Fu et al. (2019). Generally, the CMAQ model faithfully reproduced the observed concentrations of $NO_2$, $SO_2$, CO, $PM_{2.5}$ and $O_3$ with low biases during both peirods. The evaluation results suggested reasonable estimations of the anthropogenic emissions for the year 2020 and during the lockdown.

As we focused on the O$_3$ changes in NC, CC, and SC regions, we futher evaluated the modeling performace in simulating the variations of O$_3$ and NO$_2$ for these regions. Fig. S1 shows the time series of simulated and observed O$_3$ and NO$_2$ mixing ratios. The magnitude and variation of the observed NO$_2$ mixing ratios for these three regions were all well captured by the CMAQ model. The observed O$_3$ mixing ratios for three regions were also reasonably reproduced. Both the simulation and observation showed an O$_3$ increase in NC and CC but a decrease in SC (also shown in Fig. 4). However, the O$_3$ mixing ratio in NC during the lockdown was underestimated, probably due to the uncertainties in meteorological simulation. The O$_3$ mixing ratio in SC was generally overestimated during the simulation period, which might be attributed to the influence of the overestimated O$_3$ concentrations on the ocean. Nevertheless, the model was able to faithfully capture the observed O$_3$ variations in these three regions.

Overall, despite some uncertainties, the CMAQ model performance is acceptable and can support further analysis of O$_3$ changes during the COVID-19 city lockdowns.

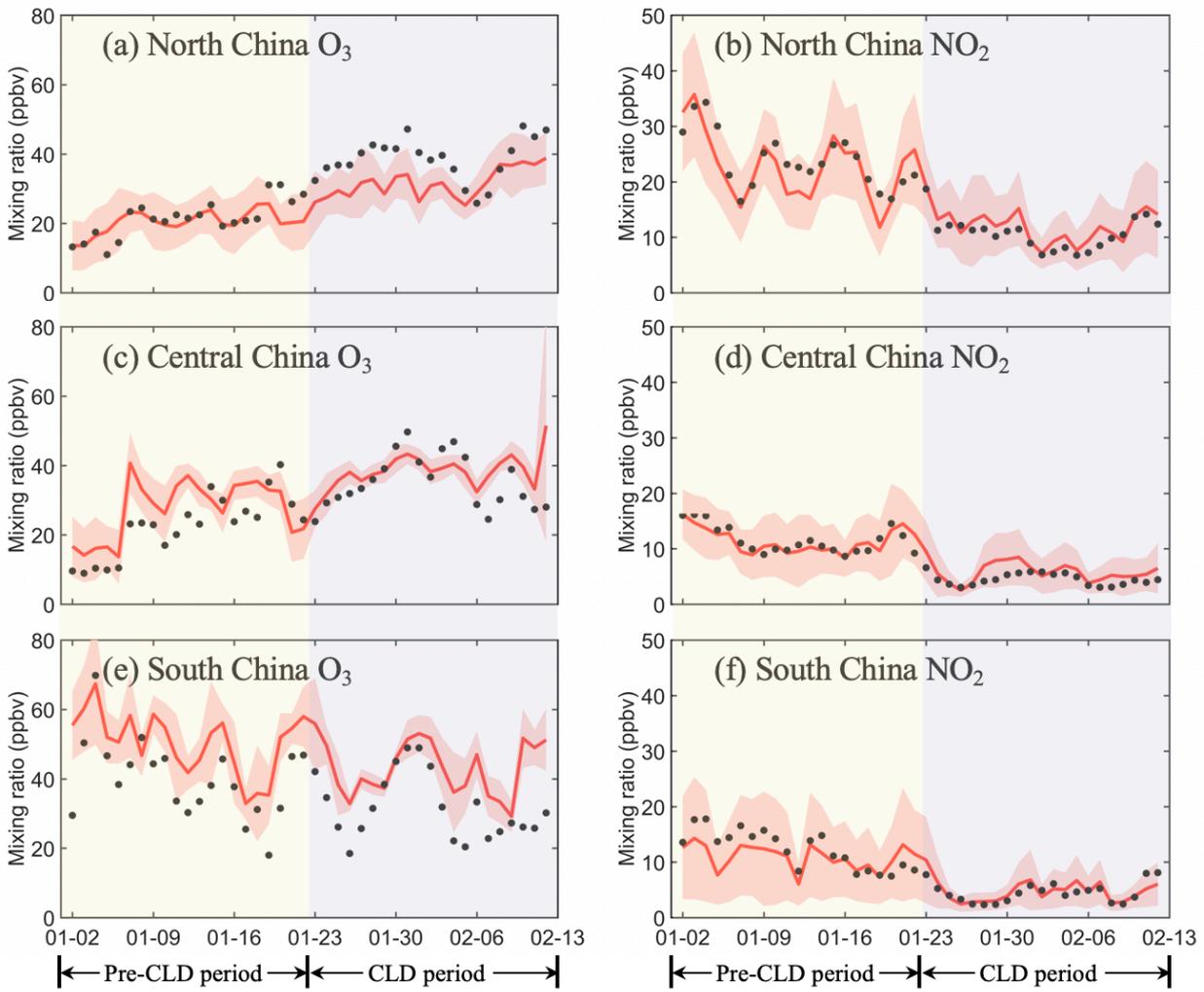

Figure S1: Time series of observed (black points) and simulated (red lines) mixing ratios of maximum daily average 8-h (MDA8) $O_3$ and $NO_2$ in North China, Central China, and South China from 2 January to 12 February 2020. The solid lines are the simulated average value and the shaded areas mark the standard deviations. The observed $NO_2$ data were adjusted based on the method proposed by Zhang et al. (2017) and Fu et al. (2019): $NO_{2\ OBS} = NO'_{2\ OBS} \times \frac{NO_{2\ SIM}}{NO_{2\ SIM} + NO_{z\ SIM} + NO_{3\ SIM}^-}$, where $NO'_{2\ OBS}$ is the measured $NO_2$ data by the catalytic conversion technique, $NO_{2\ SIM}$, $NO_{z\ SIM}$, and $NO_{3\ SIM}^-$ are the simulated data of $NO_2$, $NO_z$, and particulate nitrate, respectively, using the WRF-CMAQ model.

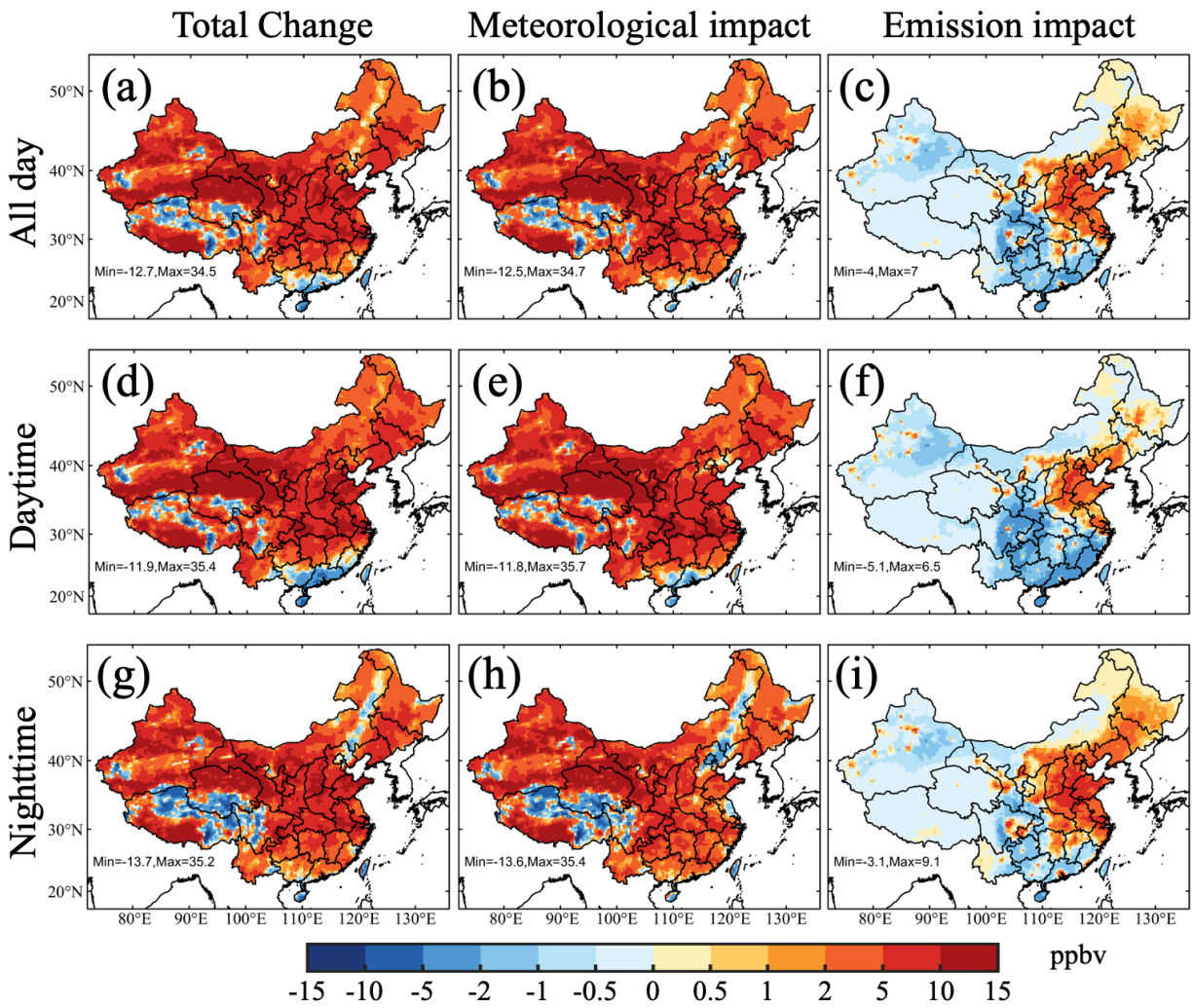

Fig. S2: Simulated changes in $O_3$ mixing ratios across China during the COVID-19 lockdown period and contributions from meteorological changes and emission reductions. (a, d, g) The simulated total $O_3$ changes for all-day average, daytime average, and nighttime average during the CLD period relative to the pre-CLD period. (b, e, h) Contribution of meteorological changes to $O_3$ for all-day average, daytime average, and nighttime average. (c, f, i) The same with (b, e, h), respectively, but for contribution of emission reductions.

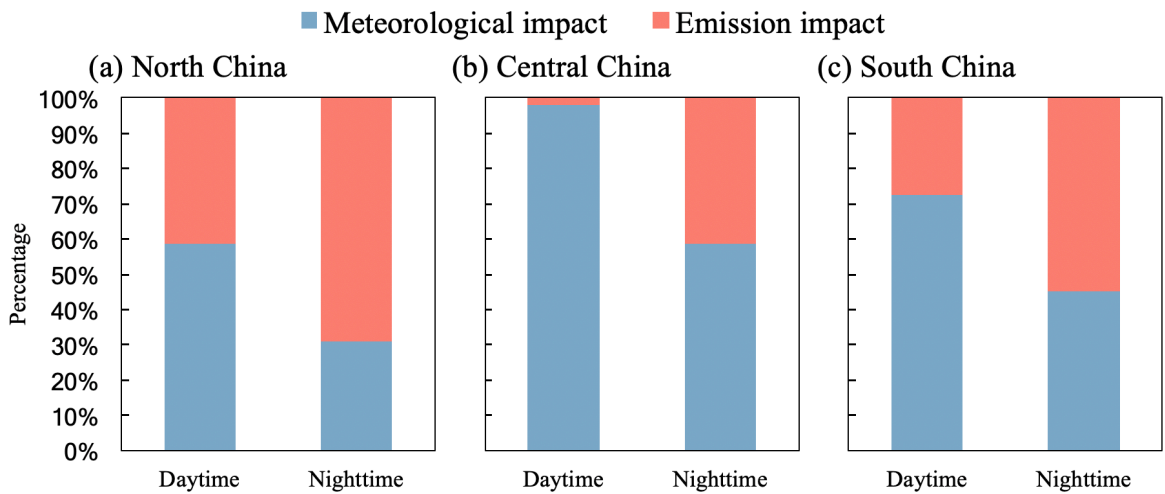

Figure S3 The contributions of meteorological changes and emission reduction to the changes in daytime and nighttime O₃ concentrations during the CLD period compared with the pre-CLD period. (a) North China; (b) Central China; (c) South China

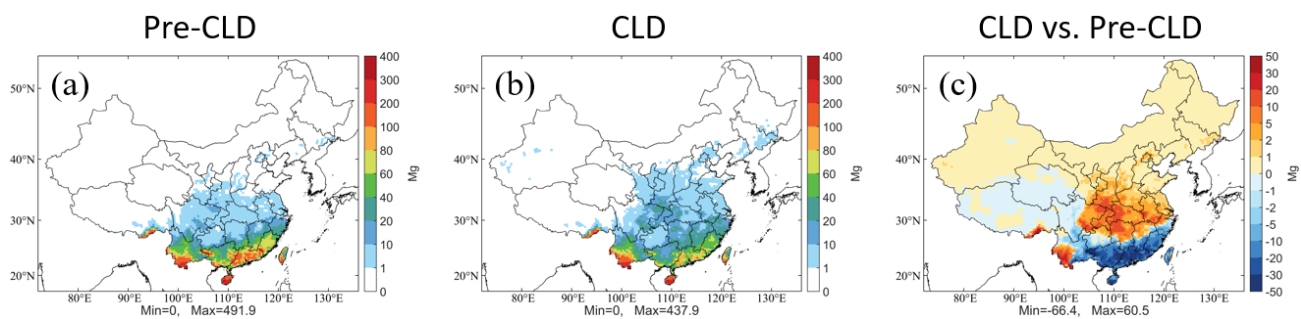

Figure S4: Biogenic isoprene emissions during the pre-CLD and CLD periods and their difference (CLD minus pre-CLD).

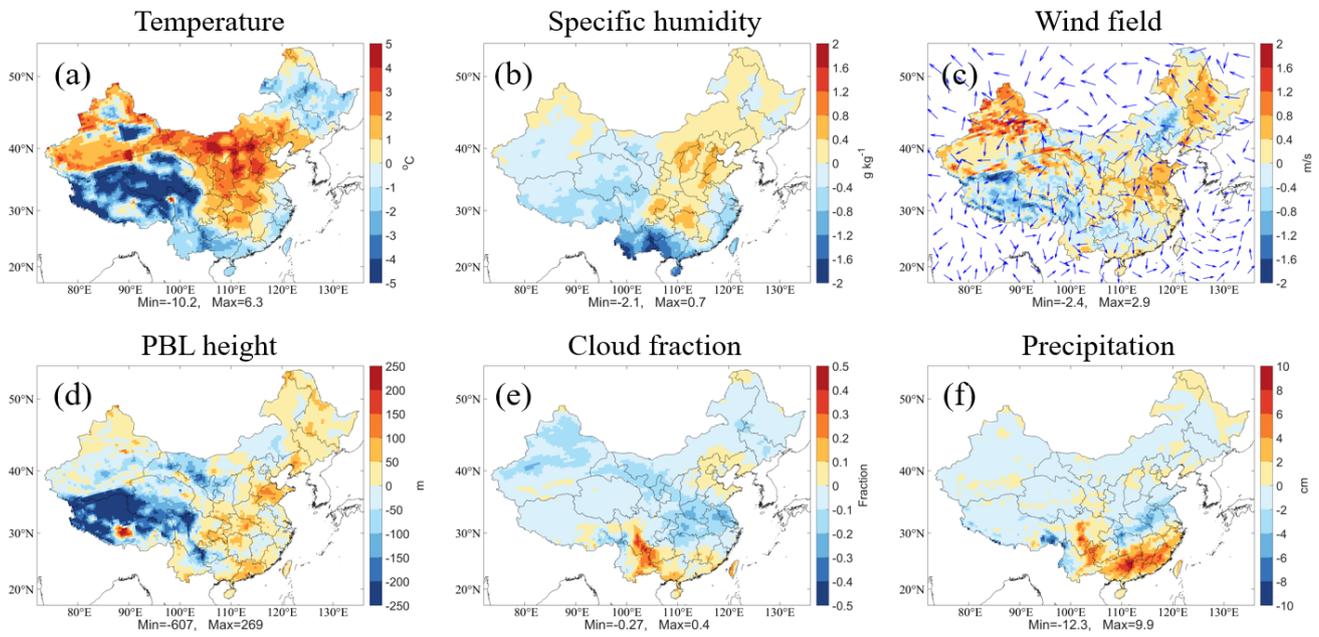

Figure S5: Model simulated changes in nighttime temperature at 2 m height, specific humidity at 2 m height, wind field at 10 m height, planetary boundary layer (PBL) height, cloud fraction, and precipitation during CLD period relative to pre-CLD period. In panel (c), the shaded color and vector represent the wind speed and wind direction, respectively.

Table S1: Scaling factors applied to different economic sectors in order to estimate the anthropogenic emissions of China for the year 2020 based on the 2017 MEIC emission inventory.

| Emitted species | Power plants | Industry | Residence | Transportation |
| --- | --- | --- | --- | --- |
| $NO_x$ | -35% | - | - | - |
| $SO_2$ | -40% | -40% | -25% | - |
| PM | - | -20% | -30% | - |

Table S2: Evaluation results of the air pollutants across China for the pre-CLD (2-22 January 2020) and CLD (23 January-12 February 2020) periods. OBS is mean observation; SIM is mean simulation; MB is mean bias; MAGE is mean absolute gross error; RMSE is root mean square error; IOA is index of agreement; $r$ is correlation coefficient; OBS, SIM, MB, MAGE, and RMSE have the same units as given in the first column, while IOA and $r$ have no unit.

| Species | Period | OBS | SIM | MB | MAGE | RMSE | IOA | $r$ |
| --- | --- | --- | --- | --- | --- | --- | --- | --- |
| $SO_2$ (ppbv) | Pre-CLD | 4.9 | 5.4 | 0.5 | 3.8 | 4.5 | 0.79 | 0.35 |
|  | CLD | 4.1 | 4.1 | 0.0 | 3.1 | 3.8 | 0.77 | 0.33 |
| $NO_2$[a] (ppbv) | Pre-CLD | 14.7 | 12.5 | -2.2 | 5.1 | 6.0 | 0.90 | 0.49 |
|  | CLD | 6.6 | 6.7 | 0.1 | 3.1 | 3.7 | 0.89 | 0.58 |
| CO (ppmv) | Pre-CLD | 0.94 | 0.66 | -0.28 | 0.42 | 0.48 | 0.88 | 0.40 |
|  | CLD | 0.75 | 0.51 | -0.24 | 0.34 | 0.40 | 0.88 | 0.40 |
| MDA8 $O_3$ (ppbv) | Pre-CLD | 26.2 | 30.1 | 3.9 | 10.5 | 13.0 | 0.94 | 0.35 |
|  | CLD | 36.0 | 37.6 | 1.6 | 10.5 | 12.9 | 0.97 | 0.38 |
| $PM_{2.5}$ (μg/m³) | Pre-CLD | 69.3 | 71.8 | 2.4 | 33.0 | 41.6 | 0.90 | 0.51 |
|  | CLD | 55.3 | 52.7 | -2.6 | 25.9 | 34.2 | 0.90 | 0.55 |

a The observed $NO_2$ data were adjusted based on the method proposed by Zhang et al. (2017) and Fu et al. (2019): $NO_{2\ OBS} = NO'_{2\ OBS} \times \frac{NO_{2\ SIM}}{NO_{2\ SIM} + NO_{z\ SIM} + NO_{3\ SIM}^-}$, where $NO'_{2\ OBS}$ is the measured $NO_2$ data by the catalytic conversion technique, $NO_{2\ SIM}$, $NO_{z\ SIM}$, and $NO_{3\ SIM}^-$ are the simulated data of $NO_2$, $NO_z$, and particulate nitrate, respectively, using the WRF-CMAQ model.